\documentclass[submission,copyright,creativecommons]{eptcs}
\providecommand{\event}{PxTP 2021} 
\usepackage{underscore}           
\usepackage{alltt}
\usepackage{tgcursor}

\usepackage{xypic}
\xyoption{curve}
\usepackage{alltt}

\usepackage{pgf,tikz}
\usepackage{wrapfig}
\xyoption{curve}
\usepackage{alltt}

\usepackage{pgf,tikz}
\usepackage{wrapfig}

\usepackage[counterclockwise]{rotating}
\usepackage[english]{varioref}
\usepackage{url}
\usepackage{xypic}
\xyoption{all}
\usepackage{varioref}

\usetikzlibrary{knots,calc,shapes.geometric}
\usetikzlibrary{arrows}
\definecolor{ttzzqq}{rgb}{0.2,0.6,0.0} 
\definecolor{ffqqqq}{rgb}{1.0,0.0,0.0} 
\definecolor{qqqqff}{rgb}{0.0,0.0,1.0} 
\definecolor{ffxfqq}{rgb}{1.0,0.4980392156862745,0.0} 
\definecolor{xfqqff}{rgb}{0.4980392156862745,0.0,1.0} 
\definecolor{zzttqq}{rgb}{0.6,0.2,0.0} 
\definecolor{ttqqqq}{rgb}{0.2,0.0,0.0}
\definecolor{mygreen}{RGB}{34,120,15}

\title{Integrating an Automated Prover for Projective
  Geometry as a New Tactic in the Coq Proof Assistant}
\author{Nicolas Magaud
\institute{ICube UMR 7357 CNRS - Université de Strasbourg}
\email{magaud@unistra.fr}
}
\def\titlerunning{Integrating an Automated Prover for Projective
  Geometry as a New Tactic in the Coq Proof Assistant}
\def\authorrunning{N. Magaud}
\begin{document}
\maketitle

\begin{abstract}
Recently, we developed an automated theorem prover for projective
incidence geometry. This prover, based on a combinatorial approach
using matroids, proceeds by saturation using the matroid rules.  It is
designed as an independent tool, implemented in C, which takes a geometric configuration as input and
produces as output some Coq proof scripts: the statement of the
expected theorem, a proof script proving the theorem and possibly some
auxiliary lemmas.  In this document, we show how to embed such an
external tool as a plugin in Coq so that it can be used as a simple
tactic.  

\end{abstract}

\section{Introduction}

The Coq proof assistant~\cite{coqmanual, BC04} is a generic theorem prover, with huge
capabilities. One of its main strengths is that it allows carrying out
proofs either interactively or under some assumptions automatically.  
Its standard tactics  allow to solve goals in some well-understood
fragments of its underlying logic, e.g. first order logic using the
tactic \texttt{firstorder} or linear arithmetic using the tactic \texttt{lia}.
When the proofs get more technical or are outside of the scope of
these automatic tactics, the user can take control and thus we are not
limited any more by the power of automation.  Contrary to what happens
with SMT provers, like Z3~\cite{MouraB08} or
Vampire~\cite{vampire}, which either succeed or fail on the goal, in
Coq, the user can perform some proof steps interactively and then rely
again on automated tools to solve the goal.

However, it requires a lot of expertise to be able to
develop some new tactics in Coq to help the user proving theorems more
easily. Most tactics
are written using the Ltac tactic language~\cite{DBLP:conf/lpar/Delahaye00}, but writing tactics
sometimes also requires to implement some features in OCaml in a Coq plugin.

In the context of geometry, we developed an independent theorem
prover, implemented in C, which handles first-order statements. It takes as input a geometric
configuration, written as plain text, and returns a Coq proof script,
ready to be verified (type-checked) by Coq.  We choose to proceed that
way in order to make the development as easy as possible for a
non-expert in Coq internals. The only requirements are to know the
specification and the tactic languages of Coq, and thus to be able to
produce some Coq files (actually a simple \texttt{.v} file).

In this article, we show how to integrate such an automated prover as a tactic in
Coq, without modifying the interface of the automated prover.

In Sect.~\ref{sect:prover}, we briefly present the automated prover, which we
shall see as a blackbox in the rest of this document. In Sect.~\ref{sect:input},
we show how to translate a Coq goal into the input language of our
prover. In Sect.~\ref{sect:output}, we investigate how to use the Coq script
produced by our prover to solve the initial Coq goal. In Sect.~\ref{sect:discussion}, we discuss our
implementation choices and explain how it could be extended to other
external provers.

\section{The automated prover}\label{sect:prover}
We work in the framework of projective incidence geometry, which is
one of the simplest theory that can be used to capture most aspects of
geometry. It is based on an incidence relation between points and
lines. In its projective version, we assume that two coplanar lines
always intersect.  One of its main advantages is that it can be
described using only a few axioms; thus it is a well-suited framework
to try and automate proofs of theorems.  
This geometry can be described either in a synthetic way, writing
statements using the incidence relation $\in$, or in a more combinatorial
way, using the rank of a set of points~\cite{MS06}.  For instance, the property
that a line has at least three distinct points can be expressed in a
synthetic way as follows :
$$\forall  l : Line, \exists A B C : Point,
A \neq B \land B \neq C \land A \neq C \land A \in l\land B \in  l
\land C  \in l.$$
It can also be implemented as the following statement using the matroid 
based description using ranks : 
$$
\forall A B : Point,  rk\{A, B\} = 2~\Rightarrow~\exists C : Point, rk\{A, B, C\} = rk\{B, C\} = rk\{A, C\} = 2.
$$

\begin{figure}
\begin{center}
 \scalebox{2}{\begin{tikzpicture}[line cap=round,line join=round,>=triangle 45,x=1.0cm,y=1.0cm]
    \tikzstyle{every node}=[font=\tiny]
    \clip(-1.0,-1.0) rectangle (1.0,1.0);
    \fill[color=white,fill=white,fill opacity=0.1] (-0.6770714035101904,-0.4325980198719555) -- (0.34110966716268676,0.6899787854448113) -- (0.7603606962632832,-0.708538687100658) -- cycle;
    \fill[color=white,fill=white,fill opacity=0.1] (-0.6770714035101904,-0.4325980198719555) -- (-0.12848984015185505,0.17223048911171895) -- (0.7603606962632832,-0.708538687100658) -- cycle;
    \draw (-0.6770714035101904,-0.4325980198719555)-- (0.34110966716268676,0.6899787854448113);
    \draw (0.34110966716268676,0.6899787854448113)-- (0.7603606962632832,-0.708538687100658);
    \draw (-0.12848984015185505,0.17223048911171895)-- (0.7603606962632832,-0.708538687100658);
    \draw [color=zzttqq] (-0.6770714035101904,-0.4325980198719555)-- (0.34110966716268676,0.6899787854448113);
    \draw [color=zzttqq] (0.34110966716268676,0.6899787854448113)-- (0.7603606962632832,-0.708538687100658);
    \draw [color=zzttqq] (0.7603606962632832,-0.708538687100658)-- (-0.6770714035101904,-0.4325980198719555);
    \draw [color=zzttqq] (-0.6770714035101904,-0.4325980198719555)-- (-0.12848984015185505,0.17223048911171895);
    \draw [color=zzttqq] (-0.12848984015185505,0.17223048911171895)-- (0.7603606962632832,-0.708538687100658);
    \draw [color=zzttqq] (0.7603606962632832,-0.708538687100658)-- (-0.6770714035101904,-0.4325980198719555);
    \draw (0.7603606962632832,-0.708538687100658)-- (-0.6770714035101904,-0.4325980198719555);
    \begin{scriptsize}
    \draw [fill=black] (-0.6770714035101904,-0.4325980198719555) circle (1pt);
    \draw[color=black] (-0.8554417739004563,-0.4024269894667369) node {$A$};
    \draw [fill=black] (0.34110966716268676,0.6899787854448113) circle (1pt);
    \draw[color=black] (0.33795740378599054,0.864211946506821) node {$D$};
    \draw [fill=black] (0.7603606962632832,-0.708538687100658) circle (1pt);
    \draw[color=black] (0.9018833300302452,-0.7083676566954394) node {$B$};
    \draw [fill=black] (-0.12848984015185505,0.17223048911171895) circle (1pt);
    \draw[color=black] (-0.28064341960192262,0.2321681329971913) node {$C$};
    \end{scriptsize}
\end{tikzpicture}}
\end{center}
\caption{A simple geometric configuration illustrating the statement
  of Fig.~\ref{fig:input}}\label{fig:triangle}
\end{figure}
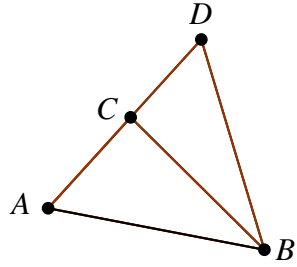

In the combinatorial approach, we exclusively deal with points and geometric reasoning is replaced by some computations relying on the
combinatorial properties of the underlying matroid on which the rank
properties are based \cite{MS06}.  
Concretely, the rank of a set of two distinct points A and B is equal
to 2. The rank of a set of three collinear points A, B, and C is also equal
to 2. The rank of the whole plane in a two dimensional setting is
3. In a space of dimension $n\ge 3$, the maximum rank of a set of
points is $n + 1$. In the case $n = 3$, a set of points whose rank is 4 is not a plane
and actually captures the whole space.

The above two approaches are shown to be equivalent \cite{BMS18a} and
the combinatorial one 
can be successfully used to automatically prove some emblematic
theorems of 3D projective incidence geometry~\cite{Braun2019}.  Among
them we can cite Desargues' theorem and Dandelin-Gallucci's theorem~\cite{BMS21}. 
The automated prover, named Bip for \textsl{matroid \textbf{B}ased \textbf{I}ncidence \textbf{P}rover},  is designed to prove equality between ranks of
various sets of points. It is based on rank interval computations. For
each subset of the powerset of the geometric configuration, we define
the minimum and the maximum rank (in the worst case, when no
information is known, the rank of each non-empty subset is between 1
and 4). We then use the matroid axioms, which specify the rank
function, and we reformulate them as rewrite rules to incrementally reduce
the size of the interval for each subset. This is achieved using a
saturation algorithm, which is run on a valuated graph implementing the inclusion lattice of the point
powerset, labeled by the minimum and maximum rank.  Once the saturation
graph is built, it is traversed to build a Coq proof script which
actually proves the statement at stake. Technical details about the
implementation can be found in \cite{Braun2019}  and the current implementation of the prover as well as several examples of
applications are available in the git
repository: \url{https://github.com/pascalschreck/MatroidIncidenceProver}.

In the  next sections, we show how to integrate such a tool so that it can
be simply used as a tactic in Coq.  

\section{Translating a Coq statement into the input language of the prover}\label{sect:input}
Provided that we define the theory of ranks in Coq (see Appendix \ref{sect:rank}), as we first did in
\cite{MNS12}, we can state geometric properties as Coq lemmas using
equalities on the ranks of some sets of points.

A Coq statement featuring
some rank properties can easily be translated into an input file for the automated prover.
It simply consists in traversing the context and the goal, harvesting
all variables of type \texttt{Point} and all statements of the form
rk(e)=k. All other statements are simply dropped, as they could not be
used by the automatic prover yet.  Fig.~\ref{fig:input} shows
the correspondence between the Coq statement of a simple theorem in 
projective geometry,
depicted  in Fig.~\ref{fig:triangle}, 
and its translation into the input langage of our
automated prover.

\begin{figure}
\begin{minipage}{0.65\linewidth}
\begin{alltt}
\textbf{Lemma} ex2 : 
    forall A B C D:Point,
    rk(A :: D :: B :: nil) = 3 ->
    rk(A :: C :: D :: nil) = 2 ->
    rk(C :: A :: nil) = 2 ->
    rk(C :: D :: nil) = 2 ->
    rk(A :: C :: B :: nil) = 3.
\textbf{Proof.}
\end{alltt}
\end{minipage}
 \begin{minipage}{0.34\linewidth}
\begin{alltt}
\textsl{\textbf{context}}
  \textbf{dimension} 3
  \textbf{layers} 1
\textsl{\textbf{endofcontext}}
\textsl{\textbf{layer 0}}
 \textbf{points}
A B C D 
 \textbf{hypotheses}
C D  : 2
C A  : 2
A C D  : 2
A D B  : 3
  \textbf{conclusion}
A C B  : 3
\textsl{\textbf{endoflayer}}
\textbf{conclusion}
A C B  : 3
\textsl{\textbf{end}}
\end{alltt}
\end{minipage}
\caption{The Coq statement of a simple projective geometry theorem (left) and the corresponding input
  for the automated prover (right)}\label{fig:input}
\end{figure}

This translation can be easily implemented inside a
Coq plugin, following the examples of the plugin tutorial available in
the reference manual of Coq \cite{coqmanual}. It produces a simple
text file, which can be then used as input by the Bip prover. In the
generated input of the Bip prover shown in the right part of Fig.\ref{fig:input}, layers are
a legacy tool which was used to decompose the context at stake. This
allows to build smaller proof scripts which can be easily checked by Coq. In the current
version of the prover, each deduced statement is stated as a lemma and
thus layers are not needed anymore.  Once the input file is generated, we
simply execute the external prover from Coq using the \texttt{Unix.system}
primitive of the OCaml library.  It produces a Coq file (\texttt{.v}),
with the appropriate prelude, which is ready to be verified by Coq. 

An implementation of this translation as well as the mechanisms to use
the produced proof inside Coq is available in this git repository: \url{https://github.com/magaud/projective-prover}.  It
works with Coq 8.12.2 (December 2020) and we are currently updating it
to the most recent release of Coq. 

\section{Running the (generated) Coq script and proving the goal}\label{sect:output}

Assuming the automated prover manages to produce a Coq proof script
which proves the goal at stake, we still need to load this Coq file
inside Coq and use the externally proven lemma to actually prove
the initial goal. 
Let us check what happens with our simple example. As the automated
prover reorders the points in the sets at stake (they are ordered
according to the order in which they were introduced, during the
construction of the input file), the
automatically generated statement and the initial one slightly differ.
In our case, the prover produces the following statement
in a file named \texttt{pprove\_ex2.v}\footnote{The name of the Lemma
  is automatically generated, starts from the \texttt{L} of Lemma,
  followed by all the names of the points used in the conclusion.}:
\begin{alltt}
\textbf{Lemma} LABC : forall A B C D ,
  rk(A :: C ::  nil) = 2 ->
  rk(A :: B :: D ::  nil) = 3 ->
  rk(C :: D ::  nil) = 2 ->
  rk(A :: C :: D ::  nil) = 2 ->
  rk(A :: B :: C ::  nil) = 3.
\textbf{Proof.}
\end{alltt}
Proving the initial goal \texttt{ex2} from Sect.\ref{sect:input} requires compiling
all the automatically generated code leading to the proof of lemma
\texttt{LABC} and loading it in Coq. 
Then the statement \texttt{LABC}  can be almost directly applied to
prove the statement at stake \texttt{ex2}. The slight differences
between these two statements lie in the order of the points in each
hypothesis or conclusion of the form $rk(e)=n$.
Points need to be reordered according to the order in which they are
introduced in the context. This is achieved by running the Ltac code
\texttt{solve\_using}\footnote{The code of \texttt{solve\_using} is in
the file \texttt{Ltac_utils.v}} which transforms the initial goal until its exactly
matches the automatically-proven statement.  

\begin{alltt}
Require Import pprove_ex2.
solve_using LABC.
\end{alltt}
At this point, even if we can be fairly confident that the whole
machinery works properly, we do not have a strong warranty that everything
went as planned.  
The last step of the proof is, as usual in Coq, to type-check the
built proof using the concluding command \texttt{Qed}. Once this
step is completed, we are sure that our prover actually proved the
initial statement provided by the user. 

\section{Discussion}\label{sect:discussion}

We choose a simple but efficient approach to run the prover inside 
Coq. One of the main advantages of this approach is that
it does not require the designer of the external prover to have any
understandings of the internals of Coq.  It simply requires the
developer to be knowledgeable of the way Coq scripts are written. This
is a skill every Coq user has. Such an approach thus makes every Coq
user a potential tactic writer, provided we can build an 
interface linking Coq statements to the expected inputs of such an external prover.
Indeed, the critical part of our tool is the translation of the statement from
Coq to the input language of the prover. The proof generation does not
need to be formally verified. Indeed, Coq will reject the proof
script if it does not actually prove the statement at stake (either
because statements do not coincide, or because the proof script is incorrect).  

The connection from Coq to the automated prover is fairly
straightforward. However the embedding of the automated prover into
Coq is a bit more technical. It is not fully automated
yet because calling a command (to load a compiled \texttt{.vo} file) from a tactic
requires subtle interactions with the State Transaction Machine (which allows for faster
and parallel executions of some parts of the proofs~\cite{DBLP:conf/itp/BarrasTT15})
to maintain consistency of the proof document.  
For the time being, we make the \texttt{pprove} tactic display the command and tactic to
be applied to complete solving the goal.  In the near future, we plan
to integrate these two extra steps directly into the code of the
\texttt{pprove} tactic.  

We think this simple way of embedding the prover
could be generalized to other external provers.
So far, our automated prover is running from scratch every time we
compute the proof again. We plan to add some memoization techniques
in order to avoid recomputing the saturation every time.  We could
aim at a more incremental tool where the saturation mechanism is
carried out when a new point is added to the context.  

More integrated tools to make automated proofs exist \cite{SATCoq},
designed to embed SAT and SMT solvers inside Coq and ensure that they
provide correct decision procedures.  Their approach is safer than ours, but we
argue that simply generating a Coq file is much easier to carry out by an
average programmer and can be used as a first step towards integrating
an automated tool into Coq.  
\section{Conclusions and Perspectives}
We successfully embed the external prover Bip as a simple tactic to be
used in Coq using Ltac.  We proceed in two steps: we first develop a
prover, which is completely independent from the internals of Coq.
The second step consists in connecting this external prover to Coq.  We
call the prover from Coq and retrieve the external proof script it
generates. We then simply feed this script to Coq to type-check it.   The API of Coq is not intended to make it
easy to embed some Coq proof scripts generated externally to prove a
goal. We believe it would be a nice feature to ease this process by
allowing a tactic to simply run a snippet of Coq code loaded from a file.  

In the longer run, improving our prover requires to make it able to
take synthetic geometry statements as input rather than sets of points
and ranks.  We carry on developing a way to
automatically translate synthetic geometry statements into ranks
equalities in Coq, so that the user deals with synthetic geometry,
whereas the prover deals with sets of points and their ranks in the
back-end. A draft implementation of this translation is available in the file
\texttt{translate.v} of the git repository.

Finally, we think our example describes an efficient and
straight-to-the-point approach  to embed an automated prover inside Coq and is useful as a first step towards a
more integrated embedding.

\bibliographystyle{eptcs}
\bibliography{generic}

\appendix
\section{A few lines of code specifying the rank function and its
  properties\label{sect:rank}}
We outline the context describing the rank function which is loaded
before starting the proof in Coq. The axioms describe the properties
of the rank function and are used in the automatically built Coq proof
script.  See files \texttt{basic\_matroid\_list.v} and
\texttt{basic_rank_space_list.v} for details.  
\begin{verbatim}
Parameter Point : Set.

Parameter eq_dec : forall A B : Point, {A = B} + {~ A = B}.

Definition equivlist (l l':list Point) := 
  forall x, List.In x l <-> List.In x l'.

Parameter rk : list Point -> nat.
Parameter rk_compat : 
  forall x x', equivlist x x' -> rk x = rk x'.

Global Instance rk_morph : Proper (equivlist ==> (@Logic.eq nat)) rk.

Parameter matroid1_a : forall X, rk X >= 0.
Parameter matroid1_b : forall X, rk X <= length X.
Parameter matroid2 : 
  forall X Y, incl X Y -> rk X <= rk Y.
Parameter matroid3 : 
  forall X Y, rk(X ++ Y) + rk(list_inter X Y) <= rk X + rk Y.

Parameter rk_singleton_ge : 
  forall A, rk (A :: nil)  >= 1.
Parameter rk_couple_ge : 
  forall A B, ~ A = B -> rk(A :: B :: nil) >= 2.

Parameter rk_three_points_on_lines : 
  forall A B, exists C, rk (A :: B :: C :: nil) = 2 /\ 
                        rk (B :: C :: nil) = 2 /\ 
                        rk (A :: C :: nil) = 2.

Parameter rk_inter : 
  forall A B C D, rk(A :: B :: C :: D :: nil) <= 3 -> 
  exists J : Point, rk(A :: B :: J :: nil) = 2 /\ 
                    rk (C :: D :: J :: nil) = 2.

Parameter rk_lower_dim : 
  exists A0 A1 A2 A3, rk( A0 :: A1 :: A2 :: A3 :: nil) = 4.
Parameter rk_upper_dim : 
  forall e, rk(e) <= 4.
\end{verbatim}
\section{An example of an automatically generated file\label{sect:example}}
The file \texttt{pprove\_ex2.v}, which is 163-line long including
comments, is automatically generated in the
process of proving the statement \texttt{ex2}. It is available (or can
be regenerated) from
\url{https://github.com/magaud/projective-prover/blob/master/theories/pprove_ex2.v}.
All statements, including intermediary statements have exactly the same
hypotheses. In addition, there is exactly one statement for each deduction achieved
during the saturation phase of the algorithm.  The last statement
\texttt{LABC} corresponds to the actual statement proved
automatically. This lemma is then reused to prove the initial Coq
goal: \texttt{ex2} in our example.
\begin{verbatim}
Require Import lemmas_automation_g.

Lemma LB : forall A B C D ,
rk(A :: C ::  nil) = 2 -> rk(A :: B :: D ::  nil) = 3 -> 
rk(C :: D ::  nil) = 2 -> rk(A :: C :: D ::  nil) = 2 -> 
rk(B ::  nil) = 1.
Proof. [...] Qed.

Lemma LAC : forall A B C D ,
rk(A :: C ::  nil) = 2 -> rk(A :: B :: D ::  nil) = 3 -> 
rk(C :: D ::  nil) = 2 -> rk(A :: C :: D ::  nil) = 2 -> 
rk(A :: C ::  nil) = 2.
Proof. [...] Qed.

[...]

Lemma LABC : forall A B C D ,
rk(A :: C ::  nil) = 2 -> rk(A :: B :: D ::  nil) = 3 -> 
rk(C :: D ::  nil) = 2 -> rk(A :: C :: D ::  nil) = 2 -> 
rk(A :: B :: C ::  nil) = 3.
Proof.
intros A B C D 
HACeq HABDeq HCDeq HACDeq .
assert(HABCm2 : rk(A :: B :: C :: nil) >= 2).
{
	try assert(HACeq : rk(A :: C :: nil) = 2) 
by (apply LAC with (A := A) (B := B) (C := C) (D := D) ;
try assumption).
	assert(HACmtmp : rk(A :: C :: nil) >= 2) 
by (solve_hyps_min HACeq HACm2).
	assert(Hcomp : 2 <= 2) by (repeat constructor).
	assert
  (Hincl : incl (A :: C :: nil) (A :: B :: C :: nil)) 
        by (repeat clear_all_rk;my_inO).
	assert
  (HT := rule_5 (A :: C :: nil) (A :: B :: C :: nil) 
         2 2 HACmtmp Hcomp Hincl);apply HT.
}
assert(HABCm3 : rk(A :: B :: C :: nil) >= 3).
[...]
assert(HABCM : rk(A :: B :: C ::  nil) <= 3) 
by (solve_hyps_max HABCeq HABCM3).
assert(HABCm : rk(A :: B :: C ::  nil) >= 1) 
by (solve_hyps_min HABCeq HABCm1).
intuition.
Qed.
\end{verbatim}
\end{document}